\documentclass[reprint,aps,prd,nofootinbib,onecolumn,superscriptaddress,preprintnumbers]{revtex4-2}

\usepackage[a4paper,margin=2.25cm]{geometry}
\usepackage{amsmath,amssymb}
\usepackage{xcolor}
\usepackage[bookmarks=false]{hyperref}
\usepackage{amssymb,graphicx,mathtools}

\definecolor{DeepBlue}{HTML}{1F4E79}
\hypersetup{colorlinks=true,linkcolor=DeepBlue,urlcolor=DeepBlue}

\newcommand{\D}{\Delta}
\newcommand{\rplus}{r_+}
\newcommand{\rminus}{r_-}
\newcommand{\OmH}{\Omega_{\rm H}}

\newcommand{\Sig}{\Sigma}

\newcommand{\ii}{\mathrm{i}}

\newcommand{\re}{\operatorname{Re}}
\newcommand{\im}{\operatorname{Im}}

\newcommand{\cK}{\mathcal{K}}

\newcommand{\cG}{\mathcal{G}}
\newcommand{\cP}{\mathcal{P}}

\newcommand{\mmpl}{p^{\ell}}
\newcommand{\mmpn}{p^{n}}
\newcommand{\mmpm}{p^{m}}

\def\bag{\begin{aligned}}
\def\eag{\end{aligned}}
\def\bea{\begin{eqnarray}}
\def\eea{\end{eqnarray}}
\def\ba{\begin{array}}
\def\ea{\end{array}}

\def\bc{\begin{center}}
\def\ec{\end{center}}

\def\D{\Delta}

\def\nn{\nonumber}

\def\p{\phi}

\def\t{\theta}

\begin{document}

\title{Distinct Near-Horizon Trend of Synchrotron Polarization in Kerr Spacetime}
\author{Yehui Hou$^{\star}$}
\email{yehuihou@sjtu.edu.cn}
\affiliation{Tsung-Dao Lee Institute, Shanghai Jiao-Tong University, Shanghai, 201210, P.R. China}

\author{Jiewei Huang$^{\star}$}
\email{jieweihuang@stu.pku.edu.cn}
\affiliation{Department of Physics, Peking University, No.5 Yiheyuan Rd, Beijing
100871, P.R. China}

\author{Bin Chen}
\email{chenbin1@nbu.edu.cn}
\affiliation{Institute of Fundamental Physics and Quantum Technology, \& School of Physical Science and Technology, \\Ningbo University, Ningbo, Zhejiang 315211, P. R. China}
\affiliation{School of Physics, \& Center for High Energy Physics,  Peking University, No.5 Yiheyuan Rd, Beijing 100871, P.R. China}

\begingroup
\renewcommand\thefootnote{$\star$}
\footnotetext{Co-first authors}
\endgroup

\begin{abstract}
We show that the near-horizon expansion of the linear polarization vector for synchrotron emission in a Kerr background admits a distinct analytic form. 
For emission from a stationary, axisymmetric, degenerate electromagnetic field, the leading-order polarization pattern depends only on the Kerr spin and the source polar angle, while the next-to-leading-order correction further encodes the geometric and rotational structure of the electromagnetic field. Our result extends the equatorial analysis of \cite{Hou:2024qqo} and the off-equatorial leading-order result of \cite{Chael:2026fhf}. Near-horizon polarization thus offers a potential probe of the fundamental properties of rotating black holes and of gravito-electromagnetic interactions.
\end{abstract}

\maketitle

\section{Introduction}

As a fundamental astrophysical object, a spinning black hole is characterized by the rotation of its spacetime \cite{kerr1963gravitational, misner1973gravitation}. This rotation gives rise to a variety of phenomena, most notably frame dragging \cite{Lense:1918zz, Ciufolini:2004rq}, whereby particles and fields are compelled to co-rotate and develop characteristic spiral structures \cite{Blandford:1977ds, Komissarov:2004ms, Gelles:2021kti, Ricarte:2022sxg, Ricarte:2022wpd, Chael:2023pwp, Wang:2025btn}. In the immediate vicinity of the horizon, strong frame dragging significantly twists infalling matter and electromagnetic fields, driving a wide range of accretion flows toward predominantly toroidal configurations \cite{Ricarte:2022wpd, Chen:2024jkm, Ruales:2026xjb}.

Polarimetric observations provide a particularly sensitive probe of the near-horizon environment \cite{EventHorizonTelescope:2021srq, Wielgus:2022heh, EventHorizonTelescope:2024rju}. Synchrotron radiation from an optically thin medium is linearly polarized perpendicular to the local magnetic field in the emitter frame, and is thus shaped by the near-horizon toroidal field structure. As a result, it can imprint a universal \textit{Near-Horizon Polarization} (NHP) pattern on the image plane, largely independent of the detailed properties of the flow. This behavior was first identified in \cite{Hou:2024qqo}, where a controlled near-horizon expansion of the polarization vector for equatorial emitters was carried out up to next-to-leading order in the Kerr redshift factor $\Delta$. That work also demonstrated that the same NHP structure persists in Kerr--Newman--Taub--NUT spacetime. This program was subsequently extended to Kerr--Horndeski geometry \cite{Chen:2025ysv}. More recently, the leading-order NHP was established away from the equatorial plane \cite{Chael:2026fhf}, and was shown to closely trace the time-averaged near-equatorial general relativistic magnetohydrodynamics (GRMHD) emission, underscoring its observational relevance.

Despite this progress, it remains premature to claim full universality of the NHP throughout the near-horizon region, as the global subleading corrections across the entire region are not yet understood. Although the exact polarization on the horizon itself is unobservable due to extreme redshift, the near-horizon behavior remains accessible and encodes valuable physical information. This naturally raises the question of whether the NHP persists at subleading order away from the equatorial plane, where jet-associated emission may also contribute.
In this work, we carry out a more detailed near-horizon expansion of the Kerr black hole. We show that, assuming stationarity, axisymmetry and degenerate electromagnetic field, the next-to-leading-order correction to the NHP encodes the derivative of the stream function and the field-line angular velocity in a distinct manner, while remaining independent of the emitter four-velocity. In the equatorial limit, our result reduces to that of \cite{Hou:2024qqo}. These findings point to a remarkable near-horizon structure in Kerr spacetime underlying this elegant polarization behavior.
Throughout, we adopt natural units with $G = c = 1$.

\medskip

\section{General scheme}
 
For synchrotron emission from an optically thin source, the polarization vector is determined by the local electromagnetic field and the photon wavevector \cite{1979Lightman}. The electromagnetic tensor satisfies
$4 F^{\alpha \mu} (*F)_{\mu\beta} = F^{\rho\lambda} (*F)_{\rho\lambda}\,\delta^{\alpha}_{\,\,\beta}$,
where $*$ denotes the Hodge dual. Accordingly, the linear polarization vector associated with a photon of wavevector $p^{\mu}$ can be expanded on the following covariant basis:
\bea\label{eq:synvector}
f^{\mu} \propto \xi_1\, F^{\mu\nu} p_{\nu} + \xi_2\, (*F)^{\mu\nu} p_{\nu} \,,
\eea
where $\xi_1$ and $\xi_2$ are real coefficients determined by the local emission process.
For an emitter with four-velocity $u^\mu$, the electromagnetic tensor admits the decomposition
$F^{\mu\nu} = u^{\mu} e^{\nu} - u^{\nu} e^{\mu} + \epsilon^{\mu\nu\alpha\beta} u_{\alpha} b_{\beta}$,
where $e^{\mu}$ and $b^{\mu}$ are the electric and magnetic fields measured in the emitter frame, respectively. In the magnetically dominated regime with vanishing electric field, this reduces to $F^{\mu\nu} = \epsilon^{\mu\nu\alpha\beta} u_{\alpha} b_{\beta}$, and the synchrotron polarization becomes $f^{\mu} \propto \epsilon^{\mu\nu\alpha\beta} u_{\alpha} b_{\beta} p_{\nu}$ \cite{Chen:2024jkm}. In this case, Eq.~\eqref{eq:synvector} simplifies to \cite{Chael:2026fhf}
\bea\label{eq:degenfmu}
f^{\mu} \propto F^{\mu\nu} p_{\nu} \,.
\eea
Although Eq.~\eqref{eq:synvector} does not explicitly depend on $u^\mu$, in magnetized plasma systems it is generally coupled to the electromagnetic field, so that any velocity dependence is implicitly encoded in $F^{\mu\nu}$ \cite{Komissarov:2004ms}.

To analyze polarized radiation in curved spacetimes, we adopt the Newman--Penrose (NP) tetrad $\{l,n,m,\bar{m}\}$ \cite{Newman:1961qr}, satisfying $l \cdot n = -1$, $m \cdot \bar{m} = 1$, and $l^2 = n^2 = m^2 = \bar{m}^2 = 0$, together with $l\cdot m = n \cdot m = l \cdot \bar{m} = n \cdot \bar{m} = 0$. The spacetime metric then takes the form $g_{\mu\nu} = -2l_{(\mu}n_{\nu)} + 2 m_{(\mu}\bar{m}_{\nu)}$. 
A Maxwell field is described by the following three complex scalars:
\bea
\Phi_0=F_{\ell m},\quad
\Phi_1=\frac12(F_{\ell n}+F_{\bar m m}),\quad
\Phi_2=F_{\bar m n}.
\label{eq:BHP-NP-convention}
\eea

After emission, the polarization vector is parallel transported along null geodesics, $p^{\mu}\nabla_{\mu}f^{\nu} = 0$.  
In spacetimes admitting a nontrivial valence-$(2,0)$ Killing spinor, and a hidden symmetry encoded by a Killing--Yano $2$-form $J_{\mu\nu}$, there exists a conserved Walker--Penrose constant along null geodesics \cite{Walker:1970un}, namely
\bea
\kappa
=
\left[(*J)_{\mu\nu}+ i\,J_{\mu\nu}\right]p^\mu f^\nu,
\quad
p^\alpha \nabla_\alpha \kappa=0.
\eea
For $f^\mu = F^{\mu\nu}p_{\nu}$ in a vacuum Petrov type~D spacetime, the Walker--Penrose constant, as defined in the NP tetrad, takes the form (up to an overall constant)
\bea
\kappa
\,\propto \, \Psi_2^{-\frac{1}{3}} (A-\ii B)\,,
\quad
A-\ii B
\,=\,
2\left(
\overline{\Phi_0}\,p^{\bar m}p^\ell
+\overline{\Phi_2}\,p^m p^n
+2\overline{\Phi_1}\,|p^m|^2
\right)\,,
\label{eq:AminusIB-BHP}
\eea
where $\Psi_2$ is the only nonvanishing background Weyl scalar, and $A,B$ are real functions. For subsequent polarization calculations, we consider the complex combination $A-\ii B$ and the ratio
\(Z=-A/B\), which constitutes the sole degree of freedom of the local polarization vector \cite{Himwich:2020msm}. It is related to the ratio of the real part to the imaginary part of $\kappa$ via
\bea
  z=\frac{\re\{\kappa\}}{\im\{\kappa\}}
  = \frac{\re\{\Psi_2^{-\frac{1}{3}}\}Z - \im\{\Psi_2^{-\frac{1}{3}}\}}{\im\{\Psi_2^{-\frac{1}{3}}\}Z + \re\{\Psi_2^{-\frac{1}{3}}\}}
  \,.
\label{eq:z-Z-map}
\eea
It suffices to compute the ``source-side'' variable; the remaining prefactor in the Walker--Penrose constant is incorporated via the map Eq.~\eqref{eq:z-Z-map} when evaluating the observed polarization angle.

To probe the near-horizon regime, we expand the relevant quantities in terms of a radial
variable that vanishes on the horizon. 
For a Killing horizon, a natural choice is $\Delta \propto -\chi^\mu \chi_\mu$, where $\chi^\mu$ is the horizon-generating Killing vector. Using
$\nabla_\mu \chi^2 = -2\kappa_H \chi_\mu$,
with $\kappa_H$ the surface gravity, we adopt $\Delta$ (or, equivalently, up to a nonzero multiplicative function) as the expansion variable. The expansions relevant for synchrotron emission take the form
\bea\label{eq:formalexp01}
\begin{aligned}
&  p^\ell=\D^{-1}\mmpl_{-1}+\mmpl_0+O(\D),
  \quad
  p^n=\D \mmpn_1+O(\D^2), \quad p^m=\mmpm_0+\D \mmpm_1+O(\D^2),\\
&  \Phi_i=\Phi_i^{(0)}+\D\Phi_i^{(1)}+O(\D^2)\,,\quad  A-\ii B =\D^{-1}C_{-1}+C_0+O(\D),
\end{aligned}
\eea
with two complex functions
\bea
\begin{aligned}\label{eq:Cminus1}
C_{-1} =
  2\mmpl_{-1}\overline{\mmpm_0}\,\overline{\Phi_0^{(0)}} \,, \quad C_0 =
  2\Big[
  \mmpl_{-1}\big(
    \overline{\mmpm_1}\,\overline{\Phi_0^{(0)}}
    +\overline{\mmpm_0}\,\overline{\Phi_0^{(1)}}
  \big)
  +\mmpl_0\overline{\mmpm_0}\,\overline{\Phi_0^{(0)}} 
  +2\overline{\Phi_1^{(0)}}|\mmpm_0|^2
  \Big]\,.
\end{aligned}
\eea
Note that $\Phi_2 p^m p^n \sim O(\D)$, so $\Phi_2$ does not contribute at $\mathcal{O}(\Delta)$. Writing this in multiplicative form, we obtain
\bea
\begin{aligned}
 &A-\ii B
  =
  \D^{-1}C_{-1}
  \left[
  1+\D \mathcal G+O(\D^2)
  \right], \\
& \mathcal G
=
\frac{C_0}{C_{-1}}
=
\frac{\mmpl_0}{\mmpl_{-1}}
+\frac{\overline{\mmpm_1}}{\overline{\mmpm_0}}
+\frac{\overline{\Phi_0^{(1)}}}{\overline{\Phi_0^{(0)}}}
+\frac{2\overline{\Phi_1^{(0)}}|\mmpm_0|^2}
{\mmpl_{-1}\overline{\mmpm_0}\,\overline{\Phi_0^{(0)}}}.
\label{eq:G-corrected}
\end{aligned}
\eea
The source-side variable is expanded as $Z = Z_0 + Z_1 \D + O(\D^2)$. Assuming $\im\{C_{-1}\}\neq0$, a straightforward expansion of Eq.~\eqref{eq:G-corrected} yields
\bea\label{eq:general-ImG-corrected}
\begin{aligned}
 & Z_0=\frac{\re\{C_{-1}\}}{\im\{C_{-1}\}}\,,\quad
  Z_1
  =
  \frac{\re\{C_{-1}\mathcal G\}\im\{C_{-1}\}-\re\{C_{-1}\}\im\{C_{-1}\mathcal G\}}
  {\im\{C_{-1}\}^2}
  =
  -(1+Z_0^2) \im\left\{\mathcal G\right\}\,, \\ 
& \im\left\{\mathcal G\right\} =
\im\left\{
\frac{\overline{\mmpm_1}}{\overline{\mmpm_0}}
  +\frac{\overline{\Phi_0^{(1)}}}{\overline{\Phi_0^{(0)}}}
  +2\frac{\mmpm_0}
  {\mmpl_{-1}}
\frac{\overline{\Phi_1^{(0)}}}
  {\overline{\Phi_0^{(0)}}}
\right\}.
\end{aligned}
\eea
By explicitly evaluating the coefficients in this expansion for a given spacetime geometry, one can, in principle, extract all information relevant to near-horizon polarized emission, although the procedure may be algebraically cumbersome.

\section{Expansions in Kerr}

We now specialize to a Kerr black hole, characterized by its mass $M$ and angular momentum $J$ \cite{kerr1963gravitational}. We further define the spin parameter as $a = J/M$. In Boyer--Lindquist coordinates $\left(t,r,\t,\p\right)$, 
the NP basis is realized by the Hartle--Hawking (HH) tetrad, which is regular on the future horizon \cite{chandrasekhar1998mathematical}:
\bea
\begin{aligned}\label{eq:HH}
&  \ell^\mu
  =\frac{1}{2H}\left(H,\D,0,a\right), \quad
  n^\mu =\frac{H}{\D\Sig}\left( H,-\D,0,a\right),\\
& m^\mu =\frac{1}{\sqrt2(r+\ii a\cos\theta)}
    \left(\ii a\sin\theta,0,1,\ii\csc\theta\right), \\ 
&\bar m^\mu =\frac{1}{\sqrt2(r-\ii a\cos\theta)}
    \left(-\ii a\sin\theta,0,1,-\ii\csc\theta\right).
\end{aligned}
\eea
Here the redshift factor is $\D = r^2-2Mr+a^2$, while $H = r^2+a^2$ and $\Sigma = r^2 +a^2\cos^2{\t}$. 
The horizon radii and the angular velocity of the horizon are
\bea
  r_\pm = M \pm \sqrt{M^2 - a^2}\,, \quad
  \OmH = \frac{a}{r_+^2 + a^2} = \frac{a}{2M r_+} \,.
\eea
For the non-extremal case with $a < M$, we have \(r-\rplus=d_H^{-1}\D  + O(\D^2)\), where $d_H= r_+-r_-$.  
In the extremal Kerr limit, $r_+ = r_-$ and \(r-\rplus\sim\sqrt{\D}\), which requires a modified expansion; this case is discussed in Appendix~\ref{app:extremeKerr}.

\subsection{The wave vector}

The null geodesic equations are highly tractable and reduce to first-order form in the $r$ and $\theta$ directions,
\bea\label{eq:Kerrnulleq}
&&k_{\mu} = E\left(  -1, \pm_{r} \frac{\sqrt{R(r)}}{\Delta},\pm_{\theta} \sqrt{\Theta(\theta)}, \lambda \right) \,, \\
&&R = \cP(r)^2 - \Delta \cK  \,,\\
&&\Theta =  \eta + a^2\cos^2{\theta} -\lambda^2\cot^2{\theta} \,,
\eea  
where $\cP(r) = r^2 + a^2 -a \lambda$. Here $E$ denotes the photon energy, while $\lambda$ and $\eta$ are impact parameters associated with the conserved angular momentum and Carter constant \cite{Carter:1968rr}: $\cK = \eta+(\lambda-a)^2
          =\Theta + \left( \lambda\csc\theta-a\sin\theta \right)^2$.
Projecting Eq.~\eqref{eq:Kerrnulleq} onto the HH tetrad, the components become
\begin{align}
  p^\ell=\frac{EH}{\D\Sig}\left(\cP+\sqrt R\right),\quad
  p^n=\frac{E}{2H}\left(\cP-\sqrt R\right),\quad
  p^m=E\,\frac{\pm_{\theta}\sqrt{\Theta(\theta)}-\ii\left(\lambda\csc\theta-a\sin\theta \right)}
  {\sqrt2(r-\ii\, a \cos{\theta})}\,,
  \label{eq:momentum-exact}
\end{align}

Near the horizon, we have
\(\sqrt R\to\cP_+ = H_+-a\lambda\), where $H_+=r_+^2+a^2=2Mr_+$. When \(\cP_+\neq0\), this admits an expansion in \(\D\). 
For simplicity, we employ a conical expansion for simplicity, meaning that the near-horizon expansion is performed at fixed $\theta = \theta_s$. Since the family $\theta_s \in [-\pi,\pi]$ foliates the entire near-horizon region, this treatment is legal (see Eq.~\eqref{eq:shapeZ}).
The expansion coefficients for the photon four-momentum in Eq.~\eqref{eq:formalexp01} can then be computed as
\bea
\begin{aligned}\label{eq:waveexpansion}
&p^\ell_{-1}
=\frac{2EH_+\mathcal P_+}{\Sigma_+} \,,\quad \frac{p^\ell_0}{p^\ell_{-1}}
=
\frac{2r_+}{d_H}
\left(
 \frac{1}{\mathcal P_+}
+\frac{1}{H_+}
-\frac{1}{\Sigma_+}
\right)
-\frac{\mathcal K}{4\mathcal P_+^2} \,, \quad \frac{p^\ell_0}{p^\ell_{-1}}  \in \,\rm Reals
. \\
&p^m_0
=
\frac{E(\beta_s-\ii\nu_s)}
     {\sqrt{2}(r_+-\ii c_s)}\,,
\quad
p^m_1
=
-\frac{E(\beta_s-\ii\nu_s)}
       {\sqrt{2}\,d_H(r_+-\ii c_s)^2}\,,
\end{aligned}
\eea
where we have introduced $c_s= a\cos\theta_s$, $\Sigma_+=r_+^2+c_s^2$, $\beta_s=\pm_{\theta_s}\sqrt{\Theta(\theta_s)}$, and $\nu_s=\lambda\csc\theta_s-a\sin\theta_s$ on a conical surface labeled by $\theta_s$; the sign \(\pm_{\theta}\) is determined by the branch of the polar momentum. 
These yield the ratios
\bea
\frac{\overline{p^m_1}}{\overline{p^m_0}}
=
-\frac{1}{d_H(r_++\ii c_s)}
= -\frac{r_+-\ii c_s}{d_H\Sigma_+} \,, \quad
\frac{p^m_0}{p^\ell_{-1}}
=
\frac{(\beta_s-\ii\nu_s)(r_++\ii c_s)}
     {2\sqrt{2}\,H_+\mathcal P_+}
\eea
Substituting into Eq.~\eqref{eq:general-ImG-corrected}, we obtain
\bea
\begin{aligned}
\im\{\mathcal G\}
={}&
\frac{c_s}
{(r_+-r_-)
 (r_+^2 + c_s^2)}
+
\im\left\{
\frac{\overline{\Phi_0^{(1)}}}
     {\overline{\Phi_0^{(0)}}}
+
\frac{(\beta_s-\ii\nu_s)
      (\rplus+\ii \, c_s)}
     {\sqrt{2}\,H_+\mathcal P_+}
\frac{\overline{\Phi_1^{(0)}}}
     {\overline{\Phi_0^{(0)}}
}
\right\}.
\end{aligned}
\label{eq:ImG-general-after-p}
\eea
The first term is determined entirely by the Kerr geometry and photon kinematics, whereas the remaining terms depend on the near-horizon expansion of $F^{\mu\nu}$.
If we further define $  \mathcal Q
  =\overline{\Phi_1^{(0)}}\left(\overline{\Phi_0^{(0)}}\right)^{-1}
 =Q_R+\ii\, Q_I$, 
Eq.~\eqref{eq:ImG-general-after-p} can therefore be expressed entirely in terms of real quantities:
\bea
  \im\{\mathcal G\}
  =\frac{c_s}{d_H\Sigma_+}
  +\im\!\left\{
    \frac{\overline{\Phi_0^{(1)}}}{\overline{\Phi_0^{(0)}}}
  \right\} + \frac{
   (\beta_s \rplus+\nu_s c_s)Q_I
   +(\beta_s c_s -\nu_s \rplus)Q_R}
   {\sqrt2\,H_+\mathcal P_+}.
  \label{eq:ImG-real}
\eea

\subsection{The electromagnetic field}

We adopt the electromagnetic field definitions $E^\mu = F^{t\mu}$ and $B^{\mu} = *F^{\mu t}$, which are not the physical field components themselves, but are directly related to those measured by a normal observer \cite{thorne1986black}.  
One then obtains
\begin{align}
  F_{\theta\phi} &= \Sig \sin\theta\, B^r, \quad  F_{\phi r} = \Sig \sin\theta\, B^\theta, \quad  F_{r\theta} = \Sig \sin\theta\, B^\phi, \quad F_{t\phi} = -\D \sin^2\theta\, E^\phi. \\
  F_{tr}  &= -\frac{\Sig}{\Pi}
  \left(\Sig E^r + 2Mar \sin\theta\, B^\theta \right), \quad
  F_{t\theta} = \frac{\Sig}{\Pi}
  \left(2Mar \sin\theta\, B^r - \D \Sig E^\theta \right)\,, 
\end{align}
where $\Pi = H^2 -a^2\D\sin^2{\theta}$. Substituting into the HH tetrad Eq.~\eqref{eq:HH}, we obtain
\bea  \label{eq:Phi0-BL}
 && \Phi_0 =
  \frac{\sin\theta\,(r-\ii a\cos\theta)}{2\sqrt{2}\,H}
  \left[ -\frac{\Sig\D}{\Pi}
  \left(aB^r+H\csc\theta\,E^\theta\right)
  +\D B^\phi +  \ii \frac{\Sig\D}{\Pi}
  \left(aE^r-H\csc\theta\,B^\theta\right)
  -\D E^\phi \right], \\
 && \Phi_1
  =
  \frac{\Sig}{2\Pi}
  \left[
    HE^r-a\D\sin\theta\,B^\theta
    +\ii\left(HB^r+a\D\sin\theta\,E^\theta\right)
  \right]\,,
  \label{eq:Phi1bar-BL} \\
 && \Phi_2
  =
  \frac{H\sin\theta\,(r+\ii a\cos\theta)}{\sqrt{2}\,\Pi}
  \left[
    aB^r+H\csc\theta E^\theta+\frac{\Pi}{\Sig}B^\phi
    +\ii\left(
      -aE^r+H\csc\theta B^\theta-\frac{\Pi}{\Sig}E^\phi
    \right)
  \right]\,.
\eea
We consider a stationary, axisymmetric, degenerate electromagnetic field. 
In this case, the electromagnetic tensor can be expressed in terms of three scalar functions \cite{thorne1986black}:
\bea\label{eq:degenF}
F
=
\frac{\Sigma \,I(r,\t)}
     {2\pi\Delta\sin\theta}
\,dr\wedge d\theta
\,+\,
d\psi\wedge
\left[
d\phi-\Omega_F(\psi)\,dt
\right]\,,
\eea
where $\psi$ and $\Omega_F$ denote the stream function and the field-line angular velocity, respectively, both conserved along a given field line; $I$ is the poloidal current, conserved only in the force-free limit \cite{Gralla:2014yja}. In general, these quantities are coupled to the plasma flow (particularly $I$), which in ideal GRMHD \cite{camenzind1986hydromagnetic, takahashi1990magnetohydrodynamic} satisfies a nonlinear Grad--Shafranov equation \cite{Gralla:2014yja, Huang:2020lvl, Shen:2026aye}.
The magnetic field components are given by 
\bea
  B^r=\frac{\psi_{,\theta}}{\Sig\sin\theta}\,,
  \quad
  B^\theta=-\frac{\psi_{,r}}{\Sig\sin\theta}\,,
  \quad
  B^\phi=\frac{I}{2\pi\D\sin^2\theta}\,,
\eea
where $\psi_{,r} = \partial_r \psi$ and $\psi_{,\theta} = \partial_\theta \psi$. The electric field components follow analogously, with $E^\phi=0$.
The finiteness of the energy flux into the event horizon is ensured by the Znajek regularity condition \cite{Blandford:1977ds}:
\bea\label{eq:Znajek}
I(r_+,\theta)
=
\frac{4\pi M r_+\sin\theta}{\Sigma_+}
\left(\Omega_F-\Omega_H\right) \psi_{,\theta}(r_+,\theta).
\eea
Substituting these field components into \(\Phi_0=F_{\ell m}\), and using
Eq.~\eqref{eq:Znajek}, yields the leading-order term at the horizon,
\bea
  \Phi_0^{(0)}
  =
  \frac{(\rplus-\ii \,a\cos{\theta})W_+(\psi_{,\theta})_+}
  {\sqrt2\,H_+\Sig_+}\,,
  \label{eq:Phi0-leading}
\eea
where $W_+=H_+\left(\Omega_+-\Omega_H\right)$, with $\Omega_+ = \Omega_F(\psi_+)$.
The subleading terms can also be computed, conveniently expanded about a conical surface with $\theta = \theta_s$; however, for our purposes only the ratios appearing in Eq.~\eqref{eq:ImG-real} are needed. After a sequence of intermediate steps, which we omit for brevity, we obtain
\bea\label{eq:Phi0-ratio-imag} 
&& \im\left\{\frac{\overline{\Phi_0^{(1)}}}{\overline{\Phi_0^{(0)}}}\right\}
  =
  -\frac{c_s}{d_H\Sig_+}
  -\frac{q_\psi D_+}{2}\,, \quad
\im
  \left\{
    \frac{\overline{\Phi_0^{(1)}}}{\overline{\Phi_0^{(0)}}}
    +
    \frac{\overline{p^m_1}}{\overline{p^m_0}}
  \right\}
  =
  -\frac{q_\psi D_+}{2}. \\
&&  D_+=\frac{1-a\Omega_+\sin^2\theta_s}{\sin\theta_s W_+} \,, \quad q_\psi=\left.\frac{\psi_{,r}}{\psi_{,\theta}}\right|_{r_+,\theta_s}\,.
 \label{eq:phase-cancellation-core}
\eea
The same set of field components further yields
\bea
  &&\frac{\overline{\Phi_1^{(0)}}}{\overline{\Phi_0^{(0)}}}
  =
  -\frac{H_+}{\sqrt2(\rplus+\ii  c_s)}
  \left(q_\psi+\ii D_+\right)\,, \quad \frac{2p^m_0\overline{\Phi_1^{(0)}}}
       {p^\ell_{-1}\overline{\Phi_0^{(0)}}}
  =
  -\frac{(\beta_s-\ii\nu_s)(q_\psi+\ii D_+)}
        {2\cP_+}\,, \\
&&\im\left\{
  \frac{2p^m_0\overline{\Phi_1^{(0)}}}
       {p^\ell_{-1}\overline{\Phi_0^{(0)}}}\right\}
  =
  \frac{\nu_s q_\psi-\beta_s D_+}{2\cP_+} \,.
  \label{eq:Phi1bar-ratio-core}
\eea

\section{Kerr NHP Formula}

\subsection{Leading-order term}

We now evaluate the NHP expressions order by order.
Before substituting the Kerr horizon data, let us translate the
Laurent expansion of \(A-\ii B\) into the
expansion of \(\kappa\).  The Weyl scalar in Kerr takes $\Psi_2 = -M\left( r - i a\cos{\theta}\right)^{-3}$. The ratio $z$ follows from Eq.~\eqref{eq:z-Z-map}:
\bea
  z=\frac{a\cos\theta_s+r Z}{r-a\cos\theta_s\,Z}\,,
\eea
or equivalently
\(Z_0=(r z_0-a\cos\theta_s)(r+a\cos\theta_s\,z_0)^{-1}\). 
Within the expanison $ z=z_0+\D z_1+O(\D^2)$, the next-to-leading-order coefficient is then most compactly written as
\bea
  z_1=
  \frac{(\rplus+a\cos\theta_s\,z_0)^2}{\Sig_+}\,Z_1
  -\frac{a\cos\theta_s}{d_H\Sig_+}(1+z_0^2)\,.
  \label{eq:z1-from-Z1}
\eea
The coefficients above are well defined in the nondegenerate
regime implying
\(C_{-1}\neq0\), as is made explicit in Eq.~\eqref{eq:Cminus-one-phase}.
The displayed \(z\)-chart additionally assumes
\(\rplus-a\cos\theta_s\,Z_0\neq0\); if it vanishes, one should use the
 ratio $z$, or equivalently the electric vector position angle (EVPA) itself.

Substituting Eq.~\eqref{eq:Phi0-leading} and Eq.~\eqref{eq:waveexpansion}
into Eq.~\eqref{eq:Cminus1}, we obtain
\bea
  C_{-1}
  =
  \mathcal N\,(\beta_s+\ii\nu_s),
  \quad
  \mathcal N
  =
  \frac{2E^2W_+(\psi_{,\theta})_+\cP_+}{\Sig_+^2}.
  \label{eq:Cminus-one-phase}
\eea
Under nondegenerate assumptions, \(\mathcal N\neq0\).
It then follows from Eq.~\eqref{eq:general-ImG-corrected} that
\bea
  Z_0 =
  \frac{\beta_s}{\nu_s}\,,
  \label{eq:Z0}
\eea
which is independent of the electromagnetic field. If the corresponding ray reaches a distant observer’s screen with viewing angle $\theta_o$ and screen coordinates \((\beta_o,\nu_o)\) \cite{Himwich:2020msm}, then the leading-order EVPA is
\bea\label{eq:chi0}
  \chi_+
  =
  \tan^{-1}\!\left(\frac{a\cos\theta_s}{\rplus}\right)
  +\tan^{-1}\!\left(\frac{\beta_s}{\nu_s}\right)
  -\tan^{-1}\!\left(\frac{\beta_o}{\nu_o}\right)\,,
  \label{eq:chi-plus}
\eea
recovering the key result of \cite{Chael:2026fhf} for general off-equatorial emission.

\subsection{Next-to-leading-order term}

We now evaluate the next-to-leading-order correction, which captures the physical trend of synchrotron polarization. Using \(C_{-1}\propto\beta_s+\ii\nu_s\), one directly finds from Eq.~\eqref{eq:G-corrected} that
\[
  \frac{\re\{C_{-1}(1+\D\cG)\}}
       {\im\{C_{-1}(1+\D\cG)\}}
  =
  \frac{\beta_s}{\nu_s}
  -\D\frac{\beta_s^2+\nu_s^2}{\nu_s^2}\im\{\cG\}
  +O(\D^2).
\]
From Eq.~\eqref{eq:waveexpansion}, \(p^\ell_0/p^\ell_{-1}\) is real and therefore does not contribute to \(\im\{\cG\}\). Combining Eq.~\eqref{eq:ImG-real} and Eq.~\eqref{eq:phase-cancellation-core}, we obtain
\bea
  \im\{\cG\}
  =
  -\frac{q_\psi D_+}{2}
  +\frac{\nu_s q_\psi-\beta_s D_+}{2\cP_+}
  =
  \frac{q_\psi(\nu_s-\cP_+ D_+)-\beta_s D_+}{2\cP_+}.
  \label{eq:ImG-intermediate}
\eea
Then, using the identity
\begin{align}
  \nu_s-\cP_+ D_+
 =
  \frac{(\lambda-a \sin^2{\theta_s})W_+-(H_+-a\lambda)(1-a\Omega_+\sin^2{\theta_s})}
       {\sin{\theta_s}W_+}
=
  -\frac{\Sig_+(1-\lambda\Omega_+)}{\sin{\theta_s}W_+},
  \label{eq:key-identity-core}
\end{align}
we arrive at an explicit expression for $\im\{\cG\}$, which is
\bea
  \im\{\cG\}
  =
  \frac{
    -\beta_s(1-a\Omega_+\sin^2{\theta_s})
    -q_\psi\Sig_+(1-\lambda\Omega_+)
  }
  {2\sin{\theta_s}(H_+-a\lambda)(H_+\Omega_+-a)}.
  \label{eq:ImG-core}
\eea
Substituting this into Eq.~\eqref{eq:general-ImG-corrected}, together with Eq.~\eqref{eq:Z0}, yields
\bea
  Z_1
  =
  \frac{\beta_s^2+\nu_s^2}{\nu_s^2}\,
  \frac{
    \beta_s(1-a\Omega_+\sin^2{\theta_s})
    +q_\psi\Sig_+(1-\lambda\Omega_+)
  }
  {2\sin{\theta_s}(H_+-a\lambda)(H_+\Omega_+-a)}.
  \label{eq:Z1-core}
\eea
The EVPA admits the same expansion structure as in \cite{Hou:2024qqo}.  Namely, away from the arctangent branch cut,
\bea
  \chi
  =
  \tan^{-1}z-\tan^{-1}\!\left(\frac{\beta_o}{\nu_o}\right)
  =
  \chi_+ +\D\chi_1+O(\D^2)\,,
  \label{eq:chi-expansion}
\eea
with \(\chi_+\) given in Eq.~\eqref{eq:chi0}.  Using
Eq.~\eqref{eq:z1-from-Z1}, the next-to-leading coefficient is
\bea
  \chi_1
  =
  \frac{z_1}{1+z_0^2}
  =
  \frac{(\rplus+a\cos\theta_s\,z_0)^2}
       {\Sig_+(1+z_0^2)}\,Z_1
  -\frac{a\cos\theta_s}{(\rplus-\rminus)\Sig_+} \,.
  \label{eq:chi1-general}
\eea
Using \(Z_1=-(1+Z_0^2)\im\{\mathcal G\}\) together with the relation between \(Z_0\) and \(z_0\), this is equivalently
\bea
  \chi_1
  =
  -\im\{\mathcal G\}
  -\frac{a\cos\theta_s}{(\rplus-\rminus)\Sig_+}\,.
  \label{eq:chi1-ImG}
\eea
Substituting Eq.~\eqref{eq:ImG-core}, this becomes
\bea
  \chi_1
  =
  \frac{
    \beta_s(1-a\Omega_+\sin^2{\theta_s})
    +q_\psi\Sig_+(1-\lambda\Omega_+)
  }
  {2\sin{\theta_s}(H_+-a\lambda)(H_+\Omega_+-a)}
  -\frac{a\cos\theta_s}{(\rplus-\rminus)\Sig_+}.
  \label{eq:chi1-core}
\eea
We see that the next-to-leading-order correction depends explicitly on $q_\psi$ and $\Omega_F$, which characterize the near-horizon structure of the poloidal magnetic field and the field-line angular velocity induced by both black hole rotation and accretion flows \cite{beskin2009mhd, anile2005relativistic}.

\begin{figure}[!htp]
\centering
    \includegraphics[width=0.95\textwidth]{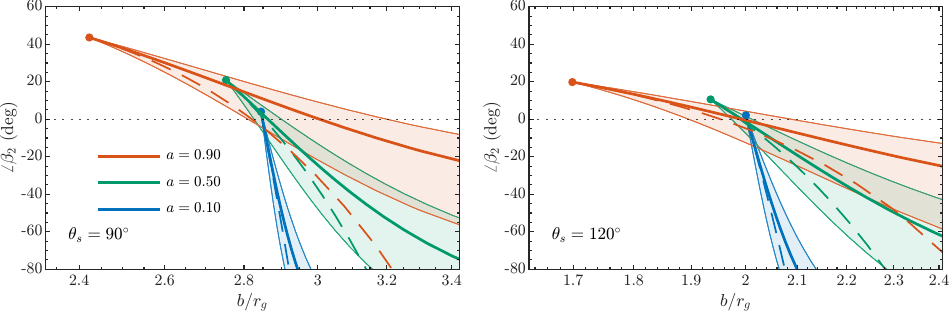}
\caption{Polarization phase $\arg\!\left(\beta_2\right)$ versus the image-plane radius $b$ for an observer at the south pole ($\theta_o = \pi$). The emission region is modeled as a cone with fixed $\theta_s$ in a split-monopole force-free magnetic field \cite{Blandford:1977ds, znajek1977black}, with $\psi = 1 - |\cos\theta_s|$. For each spin, the shaded band spans $\Omega_F \in [0,\,0.5\Omega_H]$; the solid and dashed curves correspond to $\Omega_F = 0.25\,\Omega_H$, computed exactly and to next-to-leading-order order (Eq.~\eqref{eq:chi-expansion}), respectively. Solid dots at the left end of each curve indicate the leading-order result (Eq.~\eqref{eq:chi-plus}).} 
     \label{fig:NHPvalidation}
\end{figure}

Fig.~\ref{fig:NHPvalidation} visualizes the observed polarization pattern for an observer at the south pole of the black hole. The polarization angle is encoded in the phase of the second azimuthal Fourier coefficient, $\arg\!\left(\beta_2\right)$, following the convention of \cite{Palumbo:2020flt}. Compared to the exact results obtained via full ray tracing of the polarimetric observables, the analytic expression in Eq.~\eqref{eq:chi-expansion}, including the next-to-leading-order correction, correctly reproduces the radial trend of the near-horizon polarization pattern and, sufficiently close to the horizon, further constrains the field-line rotation frequency $\Omega_F$.
This suggests that measurements of near-horizon synchrotron polarization can directly probe the black hole spin and electromagnetic field configuration, offering insight into the interaction between the black hole and the surrounding plasma.

In addition, restricting to the equatorial plane with \(\theta_s=\pi/2\), we have $\beta_s=\pm_{\theta_s}\sqrt{\eta}$, $\nu_s=\lambda-a$, and 
\bea 
Z_0^{\rm eq}=\pm_{\theta_s}\frac{\sqrt{\eta}}{\lambda-a}\,, \quad
Z_1^{\rm eq}
  =
  \frac{\eta+(\lambda-a)^2}{(\lambda-a)^2}\,
  \frac{q_\psi\rplus^2(1-\lambda\Omega_+)\pm_{\theta_s}\sqrt{\eta}(1-a\Omega_+)
  }
  {2(H_+-a\lambda)(H_+\Omega_+-a)}\,.
  \label{eq:Z1-equatorial}
\eea
If we further impose $\partial_r \psi(r,\pi/2) = 0$, corresponding to field lines reaching a turning point at the equator—a configuration commonly realized in accretion systems \cite{Komissarov:2004ms, Komissarov:2005wj}—we obtain
\bea
  Z_1^{\rm eq}(q_\psi=0)
  =
  \frac{\eta+(\lambda-a)^2}{(\lambda-a)^2}\,
  \frac{\pm_{\theta_s}\sqrt{\eta}\,(1-a\Omega_+)}
  {2(H_+-a\lambda)(H_+\Omega_+-a)}\,,
\eea
which reproduces the formula in \cite{Hou:2024qqo}. In that work, although the explicit expression was not presented, we nevertheless verified that the inclusion of $q_\psi$ does not affect $Z^{\rm eq}_0$.

\medskip

\section{Discussion}

We have derived a near-horizon expansion of the linear polarization vector for synchrotron emission in Kerr spacetime under the assumptions of stationarity, axisymmetry, and electromagnetic degeneracy (i.e., the ideal GRMHD limit). At leading order, the polarization depends only on the black hole spin and the emitter’s polar angle, consistent with \cite{Chael:2026fhf}. At next-to-leading-order order, the correction encodes the stream function and the field-line angular velocity in a specific combination, while remaining independent of the emitter’s four-velocity. This structure highlights that the leading-order polarization is purely geometric, whereas the first correction probes intrinsic properties of the electromagnetic field without direct sensitivity to plasma kinematics.

These results suggest a degree of robustness that may extend beyond the idealized setup considered here. The GRMHD simulation has indicated that the leading-order near-horizon polarization (NHP) at the equator persists in time-averaged, nearly equatorial accretion flows \cite{Wong:2025zuh, Chael:2026fhf}.
It would therefore be interesting to assess whether the NHP, including subleading corrections, remains stable in fully dynamical simulations across a broader range of spins. Time-dependent or turbulent electromagnetic fields, nonthermal emission \cite{Zhou:2025moa, Long:2026yfa}, as well as non-axisymmetric structures, could introduce departures from the analytic structure identified here or reduce its apparent universality. Clarifying the extent to which these effects modify the NHP is important for evaluating its observational utility.

Throughout the above, we fixed the polar angle to $\theta=\theta_s$, yielding an expansion on cones of constant $\theta_s$; for non-conical emission surfaces, additional chain-rule contributions from $\theta$-drift are required.
For an emitting surface $\theta=F(r)$, a point at $r=r_+ + d_H^{-1}\D$ satisfies
\bea\label{eq:shapeZ}
Z(r,\theta)=Z_0(\theta_+)+\D\left[Z_1(\theta_+)+d_H^{-1}F'(r_+)\,\partial_\theta Z_0(\theta_+)\right]+O(\D^2)\,,
\eea
with $\theta_+=F(r_+)$, indicating that the NHP formula applies to emitters of arbitrary shape, such as parabolic-like jet boundary \cite{Song:2025mhj, Gelles:2026mxg}.

Our derivation has relied on several non-degeneracy conditions to ensure the validity of the expansion. In particular, requiring a nonzero and regular leading-order coefficient 
leads to
\bea
  \Omega_H\lambda \neq 1 \,,\quad
 \Omega_+ \neq \Omega_H \,,\quad
  (\psi_{,\theta})_+ \neq 0\,,\quad
  \lambda\csc\theta_s \neq a\sin\theta_s  \,. \nn
\eea
The first condition excludes the superradiant bound, corresponding to co-rotation with the horizon and vanishing energy flux, in which case the leading contribution to the wave vector vanishes. The second condition avoids a corresponding degeneracy in the electromagnetic sector, while the third ensures that magnetic field lines thread the horizon. If \(\Omega_+=\OmH\) or \((\psi_{,\theta})_+=0\), then \(\Phi_0^{(0)}=0\), and ratios involving \(\Phi_0^{(0)}\) are not well defined. The fourth condition ensures that \(Z_0=\beta_s/\nu_s\) remains finite. When any of these conditions fails, the near-horizon expansion must be reorganized starting from the next nonvanishing order.
The extremal Kerr case discussed in Appendix~\ref{app:extremeKerr} further indicates that the near-horizon expansion and the extremal limit $a \to M$ do not strictly commute, although the leading and next-to-leading-order terms coincide with those in the non-extremal case.

Finally, our results point to a possible avenue for probing departures from Kerr. If the analytic structure identified here persists more generally, then deviations from it are expected to appear in subleading polarization signatures in alternative theories of gravity or non-Kerr rotating spacetimes \cite{Zhang:2022klr, Guo:2024bzq, Chen:2024cxi, Shi:2024bpm, Rosa:2025pqp, Angelov:2025rut, Wang:2025qpv, Yang:2025byw, Chen:2026kys}. In this sense, near-horizon polarization may offer a sensitive diagnostic of beyond-Kerr physics.

\medskip

\section*{acknowledgements}
We thank Z.Y. Zhang for helpful discussions. We also acknowledge the assistance of ChatGPT.  The work is partly supported by NSFC Grant No. 12275004, 11735001, 12588101, and 12547123.

\medskip

\appendix

\section{Near-horizon expansion for extremal Kerr}\label{app:extremeKerr}

For the extremal Kerr geometry, we set $M=a=1$ for simplicity, yielding $\rplus=\rminus=1$ and $\D=(r-1)^2$. It is convenient to introduce the variable
$$ \epsilon= r-1,\quad \D=\epsilon^2\,, $$
and the near-horizon expansions is formulated as power series in $\epsilon$. The horizon parameters reduce to $H_+=2$, $\Sig_+=1+\cos^2{\theta_s}$, $\OmH=1/2$, $\cP_+=2-\lambda$, and $W_+=2\Omega_+-1$. As in the non-extremal case, we must assume $\cP_+,W_+,(\psi_{,\theta})_+ \neq0$ to ensure regularity in the near-horizon limit.

\subsection{Factorization of leading-order terms}

Given that $R(r)=\cP(r)^2-\epsilon^2\cK$ and that $\sqrt R$ is related to the outgoing photon branch, it follows that for $\cP_+\neq0$, $\sqrt R$ is real-analytic in $\epsilon$. The components of the photon momentum are given by
$$ \begin{aligned}
  p^\ell = \epsilon^{-2}L_\ell(\epsilon)\,,\quad
  p^n = \epsilon^2L_n(\epsilon)\,, \quad
  p^m = \frac{E(\beta_s-\ii\nu_s)}{\sqrt2(1+\epsilon-\ii \cos{\theta_s})}\,,
\end{aligned} \label{eq:ext-mom-structure} $$
where $L_\ell$ and $L_n$ are real-analytic functions with $L_\ell(0)\neq0$. 

For stationary, axisymmetric, and degenerate magnetospheres, $\Phi_0$ takes the form
$$ \Phi_0 = \frac{\sin{\theta_s}(1+\epsilon-\ii \cos{\theta_s})}{2\sqrt2\,H(r)} \left[X(\epsilon)+\ii Y(\epsilon)\right]. $$
The Znajek regularity condition in Eq.~\eqref{eq:Znajek} further implies
$$ X(0)=2W_+B^r_+ = \frac{2W_+(\psi_{,\theta})_+ }{\Sig_+ \sin{\theta_s}}\neq0. \label{eq:ext-X0} $$
Furthermore, the physical requirements that $E^\phi=0$ and that $E^r$ and $B^\theta$ remain regular on the future horizon dictate that
$$ Y(\epsilon)=\epsilon^2\widetilde Y(\epsilon), \label{eq:ext-Y-order} $$
with $\widetilde Y$ being real-analytic. Therefore, the combination pertinent to Eq.~\eqref{eq:AminusIB-BHP} evaluates to
$$ \begin{aligned}
  \overline{p^m}\,\overline{\Phi_0}
  &= \frac{E(\beta_s+\ii\nu_s)}{\sqrt2(1+\epsilon+\ii \cos{\theta_s})} \frac{\sin{\theta_s}(1+\epsilon+\ii \cos{\theta_s})}{2\sqrt2\,H(r)} \left[X(\epsilon)-\ii Y(\epsilon)\right] \notag\\
  &= \frac{E\sin{\theta_s}(\beta_s+\ii\nu_s)}{4H(r)} X(\epsilon) \left[ 1-\ii\epsilon^2 \frac{\widetilde Y(\epsilon)}{X(\epsilon)} \right],
\end{aligned} \label{eq:ext-cancellation} $$
where the geometric factor $1+\epsilon+\ii \cos{\theta_s}$ cancels exactly. 

Given that $2\overline{\Phi_1}|p^m|^2 \sim \mathcal{O}(1)$, this term is parametrically suppressed by $\mathcal{O}(\epsilon^2)$ relative to the leading-order term in Eq.~\eqref{eq:AminusIB-BHP}, which scales as $2\overline{\Phi_0}\,\overline{p^m}p^\ell \sim \mathcal{O}(\epsilon^{-2})$. Additionally, we have $\overline{\Phi_2}p^m p^n \sim \mathcal{O}(\epsilon^2)$, which is $\mathcal{O}(\epsilon^4)$ relative to the leading contribution. Therefore, we can factorize this expression using a complex-analytic function $\mathcal H(\epsilon)$ and a real-analytic function $\rho(\epsilon)$ such that
$$ A-\ii B = \epsilon^{-2}(\beta_s+\ii\nu_s)\rho(\epsilon) \left[1+\epsilon^2\mathcal H(\epsilon)\right], \quad \rho(0)\neq0,\quad \rho(\epsilon)\in\mathbb R. \label{eq:ext-factorization} $$

\subsection{Absence of the $\mathcal{O}(\sqrt{\Delta})$ term and expansion structure} 

The real-valued function $\rho(\epsilon)$ strictly modulates the amplitude of $A-\ii B$, leaving its phase invariant. Let us expand $\mathcal H(\epsilon)$ as
$$ \mathcal H(\epsilon)=\mathcal H_0+\mathcal H_1\epsilon +\mathcal H_2\epsilon^2+\cdots. $$
Substituting this into Eq.~\eqref{eq:ext-factorization} yields the bracketed term $1+\epsilon^2\mathcal H(\epsilon) = 1+\epsilon^2\mathcal H_0+\epsilon^3\mathcal H_1+\mathcal{O}(\epsilon^4)$. Because $\epsilon^2 = \D$ and $\epsilon^3 = \D^{3/2}$, this explicitly demonstrates the absence of an $\mathcal{O}(\sqrt{\D})$ correction. 

Recalling that a small complex variation $\delta=u+\ii v$ to a baseline amplitude $C_0(1+\delta)$ induces a phase shift $\delta Z=-(1+Z_0^2)v+\mathcal{O}(\delta^2)$, in the extremal Kerr background, the source-side variable becomes
$$ Z = Z_0+\D Z_1^{\rm ext} +\D^{3/2}Z_{3/2}^{\rm ext} +\mathcal{O}(\D^2), \label{eq:ext-Z-expansion} $$
where we use ``ext'' to denote functions specific to the extremal limit, with the coefficients given by
$$ Z_0=\frac{\beta_s}{\nu_s},\quad Z_1^{\rm ext}=-(1+Z_0^2)\operatorname{Im}\{\mathcal H_0\},\quad Z_{3/2}^{\rm ext}=-(1+Z_0^2)\operatorname{Im}\{\mathcal H_1\}. \label{eq:ext-Z-coeffs} $$

While a generic, regular field is mathematically analytic in the radial parameter $\epsilon=r-1$, this property does not physically necessitate that $\mathcal H(\epsilon)$ be an even function. If the field smoothly extended across the horizon satisfies the parity condition $\mathcal H(\epsilon)=\mathcal H(-\epsilon)$, then $\mathcal H$ should be a function of $\epsilon^2$ (i.e., $\widehat{\mathcal H}(\D)$), restricting $Z$ to integer powers $\D^n$. However, such a symmetry is not a priori guaranteed by the stationarity, axisymmetry, the degeneracy condition, or Znajek regularity; rather, it constitutes an independent, supplementary assumption regarding the radial structure of the magnetosphere.

\subsection{Expansion coefficients at $\mathcal{O}(\D)$}

The imaginary part of $\mathcal H_0$ receives two distinct contributions. From Eq.~\eqref{eq:ext-cancellation}, the first contribution is
$$ \left(\operatorname{Im}\mathcal H_0\right)_{\Phi_0} = -\frac{q_\psi}{2}D_+^{\rm ext}, \quad D_+^{\rm ext}= \frac{1-\Omega_+\sin^2{\theta_s}}{\sin{\theta_s}(2\Omega_+-1)}. $$
The mixed term involving $\overline{\Phi_1}$ yields a contribution identical to that in the non-extremal case,
$$ \left(\operatorname{Im}\mathcal H_0\right)_{\Phi_1} = \frac{\nu_s q_\psi-\beta_s D_+^{\rm ext}}{2(2-\lambda)}. $$
Summing these gives
$$ \operatorname{Im}\mathcal H_0 = \frac{ q_\psi\left(\nu_s-(2-\lambda)D_+^{\rm ext}\right) -\beta_s D_+^{\rm ext} } {2(2-\lambda)}. $$
Using the identity $\nu_s-(2-\lambda)D_+^{\rm ext} = -\csc{\theta_s}(1+\cos^2{\theta_s})(1-\lambda\Omega_+)(2\Omega_+-1)^{-1}$ established in Eq.~\eqref{eq:key-identity-core}, we finally obtain
$$ \operatorname{Im}\mathcal H_0 = \frac{ -\beta_s(1-\Omega_+\sin^2{\theta_s}) -q_\psi(1+\cos^2{\theta_s})(1-\lambda\Omega_+) } {2\sin{\theta_s}(2-\lambda)(2\Omega_+-1)}. \label{eq:ext-ImH0} $$
Accordingly, $Z_1^{\rm ext}=-(1+Z_0^2)\operatorname{Im}\{\mathcal H_0\}$ exactly matches the non-extremal result in Eq.~\eqref{eq:Z1-core} when evaluated using extremal horizon data. However, this cannot be recovered simply by taking the limit $d_H\to0$. Instead, treating the expansion in $\epsilon=\sqrt\D$ independently is physically essential due to the non-commutativity of the near-horizon expansion and the extremal limit.

 \bibliographystyle{utphys}

\bibliography{ref}

\end{document}